\documentstyle[preprint,graphicx]{jpsj}

\def\runtitle{Real-Space Renormalization Group Study of Effects of Anisotropy 
on S=1 Random Antiferromagnetic Chain}
\def\runauthor{Shigeru {\sc Koikegami}, Synge {\sc Todo}, 
and Hajime {\sc Takayama}}

\title
{Real-Space Renormalization Group Study of Effects of Anisotropy on S=1 
Random Antiferromagnetic Chain}

\author
{Shigeru {\sc Koikegami}\footnote{E-mail: shigeru.koikegami@aist.go.jp},
Synge {\sc Todo}$^{1,2}$, and Hajime {\sc Takayama}$^{1}$}

\inst
{Japan Science and Technology Corporation\\
Nanoelectronics Research Institute, 
AIST Tsukuba Central 2, Tsukuba 305-8568\\
$^1$Institute for Solid State Physics, University of Tokyo, Kashiwa 277-8581\\
$^2$Theoretische Physik, Eidgen\"{o}ssische Technische Hochschule, CH-8093
Z\"{u}rich, Switzerland}

\recdate{today}

\abst{We investigate S=1 antiferromagnetic quantum spin chain, 
whose exchange couplings 
are strongly disordered. By the real-space renomalization group method, 
introduced by Ma, Dasgupta, and Hu, the renormalization group 
flows are analyzed numerically in a plain of the anisotoropy of the exchange
coupling vs. the staggered magnetic field. 
As the result, the Heisenberg point, which has a zero average of the 
exchange coupling anisotropy, is specified as 
the unstable fixed point against the anisotropy.}

\kword{antiferromagnetic quantum spin chain, randomness, anisotropy, 
real-space renomalization group}

\begin{document}
\sloppy
\maketitle

\section{Introduction}

As is well known, in one-dimensional S=1 antiferromagnetic Heisenberg 
(AFH) models, the ground state is quantumly disordered by the strong
quantum fluctuation.~\cite{rf:1} 
There is a finite excitation gap above their non-degenerating ground
state. Affleck {\it et al.} have exactly proved this fact 
for the model with biquadratic interactions, employing 
the valence-bond-solid (VBS) picture.~\cite{rf:3} 
The existence of the gap in the low-energy excitation makes the 
system robust against weak bond-randomness in the case with 
$J_{\rm min} \gg \Delta$, where $J_{\rm min}$ is the minimum bond
strength in all and $\Delta$ is the magnitude of the excitation
gap in the case without randomness. 

However, when the randomness becomes stronger, i.e. $J_{\rm min} \sim \Delta$, 
the situation can be changed. Two far separated spins can form a singlet
state even if they destroy the excitation gap of the bulk system. 
If the system contains an as large VBS cluster as the system size, 
the string order parameter still has a finite value. 
But if not, the string order parameter may vanish. This transition 
from quantum Griffiths (QG) phase to random singlet (RS)
phase has been investigated numerically by the exact diagonalization 
method,~\cite{rf:11} the density matrix renormalization group (DMRG) 
calculation,~\cite{rf:9} the quantum Monte Carlo (QMC) 
simulation,~\cite{rf:10} and the real-space renormalization group (RSRG) 
method.~\cite{rf:8} 

The RSRG method is convenient one to determine the explicit 
phase diagram in the ground state of a system with randomness. 
This method has been introduced by Ma, Dasgupta, and Hu to
study the S=$\frac{1}{2}$ AFH chain,~\cite{rf:2} 
and developed to being applied to other systems both in the analytical 
way and the 
numerical one.~\cite{rf:6,rf:4,rf:5,rf:8,rf:7,rf:12,rf:15,rf:16,rf:13,rf:14} 
Among these works, 
we should point out a very important one by Fisher.~\cite{rf:4}
He has fully studied the properties of 
the S=$\frac{1}{2}$ antiferromagnetic chains with various types of 
random exchange coupling in the analytical RSRG method. 
Especially, he has concluded that the Heisenberg point is nothing but 
the unstable fixed point (XXX RS fixed point), at which there occurs a 
transition from the XX RS phase to the Z (Ising) antiferromagnetic 
(ZAF) phase in a random XXZ chain.

As pointed out afterward, the RSRG method has the defect when it is applied to 
S$\geq$1 spin system.~\cite{rf:16} 
Therefore, we should take care to use it in such a way that few inadequate 
renormalization processes are involved. Some authors have created skillful 
procedures, such as the use of the effective Hamiltonian,
~\cite{rf:5,rf:15,rf:13,rf:14} and by keeping more degrees of 
freedom.~\cite{rf:8} However, these skillful procedures cannot 
be applied in a fully consistent way to models 
with uniaxial anisotropy. The uniaxial anisotropy arises 
inevitably when the renormalization process is executed in 
the S$\geq$1 quantum spin system with the magnetic field 
or the exchange anisotropy. Therefore, we should give up applying 
these skillful procedures to S=1 random AF spin chain with the anisotropy. 

In the present work we therefore adopt the 
straightforward RSRG procedure and study numerically the renomalization
group (RG) flows around the Heisenberg point under staggered magnetic field. 
The latter very much reduces occurence of the inadequate renormalization
processes, i.e., our present approach is effecient to examine the 
RG flows against the introduced anisotropy. It is found 
for the first time that the Heisenberg point, which has a zero 
average of the exchange coupling anisotropy, is specified as the 
unstable fixed point against the anisotropy. The result is in 
disagreement with the previous work by Saguia {\it et al.}~\cite{rf:12}. 
They studied almost the same model as we do, but 
without the staggered magnetic field, and concluded that the 
fixed point situates on the point with a finite average of the exchange 
anisotropy. A possible origin of the discrepancy is, we consider, the 
adequacy of the region where the decimation procedures have been 
executed.

The construction of this paper is as follows. In \S2, we introduce 
our model Hamiltonian. In \S3, we breifly show the RSRG method for 
our Hamiltonian and discuss the confidence of our numerical method. 
In \S4, the numerical results are presented, 
and in \S5 we give the summary of our analysis.

\section{Model}

Our Hamiltonian for the strongly coupled pair of spins 
${\bf S}_1$, ${\bf S}_2$ is given by
\begin{eqnarray}
 H_0 & = & J{\bf S}_1\cdot{\bf S}_2+LS_1^zS_2^z \nonumber \\
     &   &-D_1(S_1^z)^2-D_2(S_2^z)^2-h_1S_1^z-h_2S_2^z.
\end{eqnarray}
These spins are weakly coupled to the neighbors via 
\begin{eqnarray}
 {\mathcal H} & = & K_1{\bf S}_1\cdot{\bf S}_1^\prime+
 K_2{\bf S}_2\cdot{\bf S}_2^\prime+
 M_1S_1^zS_1^{\prime z}+M_2S_2^zS_2^{\prime z} \nonumber \\
 &   &  -D_3(S_1^{\prime z})^2-D_4(S_2^{\prime z})^2
 -h_3S_1^{\prime z}-h_4S_2^{\prime z} \nonumber \\
 & \equiv & {\mathcal H}^\prime
 -D_3(S_1^{\prime z})^2-D_4(S_2^{\prime z})^2
 -h_3S_1^{\prime z}-h_4S_2^{\prime z}.
\end{eqnarray}
Diagonalizing $H_0$, we obtain its eigen values, 
$\{E_k \,|\,k=0,1,\dots,8\}$, with their eigen states, 
$\{|k \rangle\, | \,k=0,1,\dots,8\}$, where we take $|0\rangle$ as 
the two-spins ground state. As long as the magnetic field is staggered, i.e.,
$h_1h_2 \leq 0$, $|0\rangle$ is always non-degenerated. 
Introducing the magnetic field, we should have both the
uniform anisotropy $L$ and the uniaxial anistropy $D$ even when 
we start the renormalization process from the state with all $L_{ij}=0$
and $D_i=0$. 

\section{Real-Space Renormalization Group Method}

In the present work we restrict our consideration to at $T=0$. 
Taking into account ${\mathcal H}$ by the perturbation 
expansion and dropping $O({\mathcal H}^3)$ terms, we can obtain 
the modified ground-state energy as
\begin{eqnarray}
E_0+\langle 0|{\mathcal H}|0 \rangle+ 
\sum_{k=1}^8 \frac{|\langle 0 | {\mathcal H} | k\rangle |^2}{E_0-E_k} 
&     & \nonumber \\
&     & \hspace*{-11.5em}\equiv E^\prime+J^\prime {\bf S}_1^\prime \cdot {\bf S}_2^\prime
 +L^\prime S_1^{\prime z}S_2^{\prime z} \nonumber \\
&        & \hspace*{-10.5em}-D_1^\prime (S_1^{\prime z})^2-
 D_2^\prime(S_2^{\prime z})^2-
 h_1^\prime S_1^{\prime z}-h_2^\prime S_2^{\prime z}.
\label{2}
\end{eqnarray}
The all non-zero matrix elements of 
\begin{eqnarray}
\langle 0|{\mathcal H}|k\rangle & \equiv & 
\langle 0|{\mathcal H}^\prime|k\rangle \nonumber \\
&   & -[D_3(S_1^{\prime z})^2+D_4(S_2^{\prime z})^2
   +h_3S_1^{\prime z}+h_4S_2^{\prime z}]\delta_{0k}, \nonumber \\
&   & k=0,1,\dots,8,
\end{eqnarray}
in eq.~(\ref{2}) should be required. 
After some algebra, we obtain
the all recursion relations of these renormalized parameters 
as summerized in Appendix \ref{11}.

Unfortunately, our recursion equations are so complicated that it would be
difficult to be solved even in approximated forms. 
Therefore, we execute the decimation 
process numerically after our predecessors 
(see Fig.~\ref{figure:5}).~\cite{rf:7,rf:12,rf:15,rf:16} 
Considering a simple case 
only with isotropic exchange couplings, we can easily find 
that this perturbative renormalization group for S$\geq$1 chain is not 
adequate.~\cite{rf:7,rf:16} In fact, 
we obtain the recursion relation for the isotropic exchange coupling of a 
spin-S chain as 
\begin{equation}
J^\prime = \frac{2S(S+1)}{3}\frac{K_1K_2}{J}
\end{equation} 
within $O({\mathcal H}^2)$, and this coeffecient 2S(S+1)/3 is larger than 
unity for S$\geq$1. However, only in the 
strongly disordered limit with staggered field, the singlet-triplet
energy-level differences are typically so larger than the nearest 
neighbor couplings that we can rarely find the inadequate 
decimation processes, in which $J^\prime > J$. 
Actually in our numerical analysis such inadequate decimation processes occur 
within 3\verb+%+ 
to the whole decimation processes. We should take into account that 
our numerical results contain the intrinsic errors within 3\verb+%+, 
and point out that our numerical conclusion is obtained within this accuracy.
\begin{fullfigure}
\begin{center}
{\rotatebox{270}{\includegraphics[width=10ex]{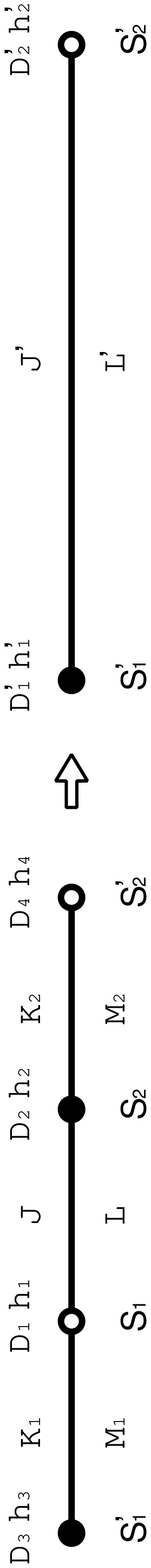}}}
\end{center}
\caption{Decimation process for our Hamiltonian. Closed and open squares
 belong to A- and B-sub-lattices respectively.}
\label{figure:5}
\end{fullfigure}

In the present work we analyze systems in the strongly disordered 
limit, i.e., the initial distribution of exchange couplings 
$\{J_{i,j}\}$ is chosen as $P(0 \le J_{i,j} \le 1) = 1$. 
We introduce the random staggered field with the distribution as 
$P(0 \leq h^A_i \leq h^{max})=1/h^{max}$ and 
$P(-h^{max} \leq h^B_i \leq0)=1/h^{max}$, where 
$h^{max}$ is varied, and $A$ and $B$ mean sub-lattices' indexes. 
Decimating until the maximum of $J_{ij}$ becomes less then 
the cutoff energy $\Omega$, a parameter which specifies the degree of 
decimation, we obtain averages of the renormalized 
parameters corresponding to $\Omega$. 
In order to obtain the reliable results, we should check 
that $J_{ij}$ is the largest energy-scale among those involved in the problem 
such as $|L_{ij}|$, $|D_i|$, and $|h_i|$, in each renormalization step. 
If, for example, we start our decimation step from the point with $\langle 
\delta \rangle > 0$, where $\delta \equiv L_{ij}/J_{ij}$ and $\langle \cdots 
\rangle$ denotes the average over the remaining bonds in each decimation 
step, already for $\Omega \sim 0.3$ we are faced to the case with $\delta 
\sim 1$ (see Fig. 3 below), where our RG rule cannot work well. 
For $\langle \delta \rangle < -0.2$, on the other hand, it turns out that 
inadequate processes occur more than 3\verb+%+. 
Therefore, we should restrict our quantitative analyses within only 
decimation steps for which $\Omega$ stays a few tenths of maximum of 
initial $J_{ij}$. For this purpose, we need to prepare such a large 
system that the averaged values can be obtained within small errors 
even if $\Omega$ stays in such a large value. 
Our system initially contains $2^{15}=32768$ spins, 
and the number of samples examined is $100$. We call the isotropic point 
with both $\langle \delta \rangle=0$ and $\langle \widetilde{h} \rangle=0$ 
the Heisenberg point, where $|\langle \widetilde{h} \rangle| \equiv 
|\langle h^A \rangle |+|\langle h^B \rangle |$. 

\section{Results}

\begin{figure}
\begin{center}
{\rotatebox{270}{\includegraphics[width=42ex]{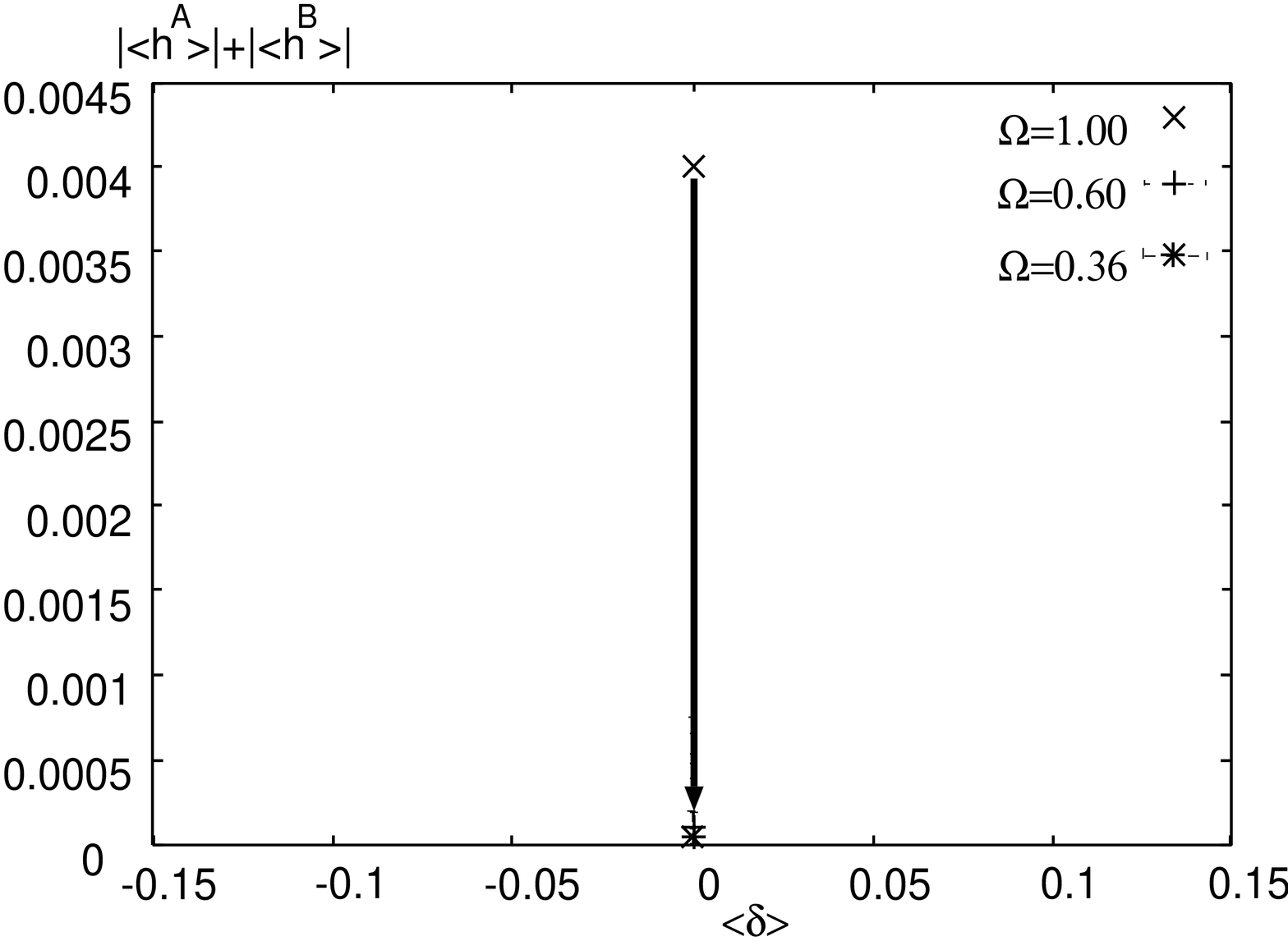}}}
\end{center}
\caption{Renormalization group flow in the decimation process starting 
from $\langle \delta \rangle=0.000$, $h^{max}=0.0040$.}
\label{figure:2}
\end{figure}

In Fig.~\ref{figure:2}, we show a typical 
RG flow in the $|\langle \widetilde{h} \rangle| - \langle \delta
\rangle$ plain. As seen in the figure, the staggered field
is an irrelevant perturbation on the line, where $\langle \delta \rangle=0$. 
We can also show that RG flows accumulate on the Heisenberg point along this
line as far as $0 < |\langle \widetilde{h} \rangle| \ll \Omega$. 

\begin{figure}
\begin{center}
{\rotatebox{270}{\includegraphics[width=40ex]{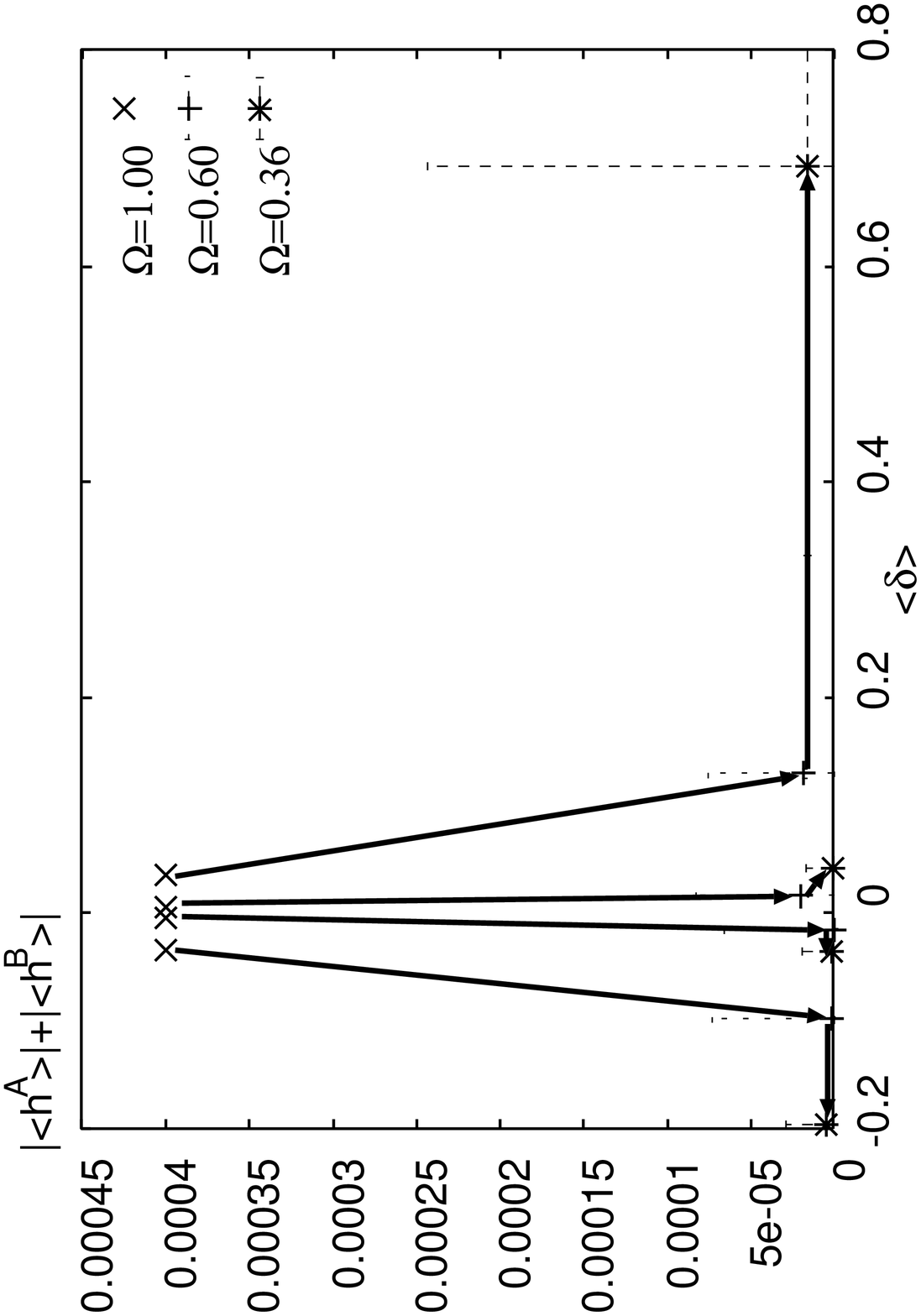}}}
\end{center}
\caption{Renormalization group flow in the decimation process starting 
from $h^{max}=0.0004$ and $\delta=\pm0.005,\pm0.035$.}
\label{figure:3}
\end{figure}

Taking the initial value of $\langle \widetilde{h} \rangle$ as 0 strictly, 
we could only obtain uncertain results on the flows near the Heisenberg 
point because the RG flows are intrinsically unstable. 
However, setting small but finite values of $\langle \widetilde{h} \rangle$ 
first, we can observe the systematic RG flows near the Heisenberg 
point as shown in Fig.\ref{figure:3}. 
In the figure, we show the RG flows under decreasing $\Omega$ from the
initial points with $ |\langle \widetilde{h} \rangle|=0.0004$ and 
$\langle \delta \rangle \neq 0$. 
The situations are completely different between $\langle \delta \rangle
> 0$ and $\langle \delta \rangle < 0$. The RG flows approach
toward the point $\langle \delta \rangle =-1$ in the area 
$\langle \delta \rangle < 0$, while they become the infinity as 
$\langle \delta \rangle \rightarrow \infty$ in the area 
$\langle \delta \rangle > 0$. These two fixed points, 
$\langle \delta \rangle=-1$ and $\langle \delta \rangle \rightarrow
\infty$ have been called the XX RS and ZAF point,
respectively, and they correspond to the different phases. 
We can also confirm that the averaged single-ion
anisotropy, $\langle D \rangle$, behaves differently between in the XX RS and
in the ZAF phases. In the XX RS phase, $\langle D \rangle < 0$, i.e.,
preferable to the low-spin state, and their absolute values are comparable to  
the cutoff energy $\Omega$. On the other hand, in the ZAF phase, 
$\langle D \rangle > 0$, i.e., preferable to the high-spin state, and
their absolute values evolve toward infinity as $\Omega$ being
decreased. All these RG flows branch off at the line 
$\langle \delta \rangle=0$. According to our more detailed
analysis, no critical line can be found more than 
$\langle \delta \rangle =0.000(3)$. 

\begin{figure}
\begin{center}
{\rotatebox{270}{\includegraphics[width=38ex]{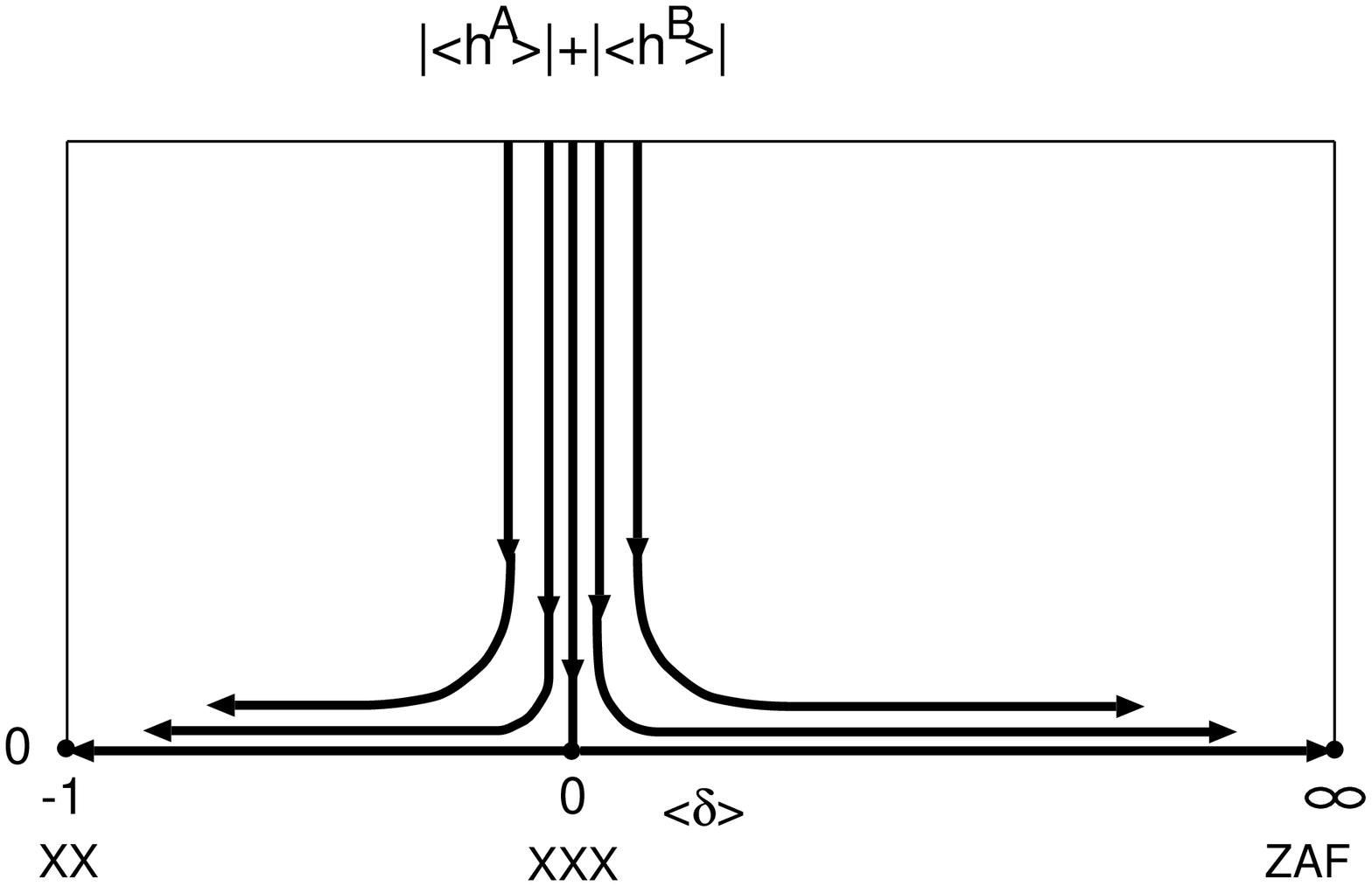}}}
\end{center}
\caption{Schematic renomalization flow diagram in the strongly
 disordered limit.}
\label{figure:4}
\end{figure}

Putting together our results discussed so far, 
we propose the schematic RG flow diagram in the strongly disordered limit 
as shown in Fig.~\ref{figure:4}, in which one sees the Heisenberg point is 
the unstable fixed point. Any definite informations about the 
feature of the Heisenberg point cannot be obtained due to the intrinsic 
numerical difficulties in our formalism. 
However, referring to the results 
by other methods,~\cite{rf:11,rf:9,rf:10,rf:8} 
we can identify this unstable fixed point as the XXX RS point. 
The feature of the Heisenberg point should be qualitatively the same as the 
Heisenberg point of the S=$\frac{1}{2}$ chain.~\cite{rf:4}

\section{Summary}

We have analyzed the strongly disordered S=1 AF quantum spin chain
by the numerical RSRG method. The RG flows around the 
Heisenberg point $\langle \delta \rangle=0$ can be
investigated by introducing the small 
staggered field, which is the irrelevant perturbation for our
renormalization process. We have carefully executed numerical 
decimation processes in such a way that few inadequate processes 
are involved. Consequently, we have obtained the reliable RG flows around 
the Heisenberg point 
within the sufficient accuracy. It is concluded that the
Heisenberg point is the unstable fixed point against anisotoropy, 
and that it turns from(to) XX RS phase to(from) ZAF phase due to
infinitesimal changes of anisotropic couplings. 

\section*{Acknowledgement}

The computation in this work was performed using SGI\,2800 at 
Supercomputer Center, Institute for Solid State Physics, 
University of Tokyo, and IBM RS/6000--SP at Tsukuba Advanced Computing
Center, National Institute of Advanced Industrial Science and Technology. 

\appendix
\section{Recursion Relations}
\label{11}

The recursion relations for the renormalized parameters in eq.~(\ref{2})
are described as:
\begin{full}
\begin{eqnarray}
E^\prime & = & E_0 
 -2\left[
 \frac{\left(A_3\right)^2+\left(B_3\right)^2}{E_3-E_0}
 +\frac{\left(A_4\right)^2+\left(B_4\right)^2}{E_4-E_0}
 +\frac{\left(A_5\right)^2+\left(B_5\right)^2}{E_5-E_0}
 +\frac{\left(A_6\right)^2+\left(B_6\right)^2}{E_6-E_0}
 \right],\nonumber \\
 J^\prime  & = & 
 2\left[-\frac{A_3B_3}{E_3-E_0}
 -\frac{A_4B_4}{E_4-E_0}
 -\frac{A_5B_5}{E_5-E_0}
 -\frac{A_6B_6}{E_6-E_0}\right], \nonumber \\
 L^\prime & = & 2\left[-\frac{A_1B_1}{E_1-E_0}
 -\frac{A_2B_2}{E_2-E_0}\right]-J^\prime, \nonumber \\
D_1^\prime & = & D_3
 +\frac{\left(A_1\right)^2}{E_1-E_0}
 +\frac{\left(A_2\right)^2}{E_2-E_0}
 -\frac{\left(A_3\right)^2}{E_3-E_0}
 -\frac{\left(A_4\right)^2}{E_4-E_0}
 -\frac{\left(A_5\right)^2}{E_5-E_0}
 -\frac{\left(A_6\right)^2}{E_6-E_0}, \nonumber \\
D_2^\prime & = & D_4
 +\frac{\left(B_1\right)^2}{E_1-E_0}
 +\frac{\left(B_2\right)^2}{E_2-E_0}
 -\frac{\left(B_3\right)^2}{E_3-E_0}
 -\frac{\left(B_4\right)^2}{E_4-E_0}
 -\frac{\left(B_5\right)^2}{E_5-E_0}
 -\frac{\left(B_6\right)^2}{E_6-E_0}, \nonumber \\
h_1^\prime & = & h_3-A_0-\frac{1}{2}\left[
 \frac{\left(A_3\right)^2}{E_3-E_0}
 +\frac{\left(A_4\right)^2}{E_4-E_0}
 -\frac{\left(A_5\right)^2}{E_5-E_0}
 -\frac{\left(A_6\right)^2}{E_6-E_0} \right], \nonumber \\
h_2^\prime & = & h_3-B_0-\frac{1}{2}\left[
 \frac{\left(B_3\right)^2}{E_3-E_0}
 +\frac{\left(B_4\right)^2}{E_4-E_0}
 -\frac{\left(B_5\right)^2}{E_5-E_0}
 -\frac{\left(B_6\right)^2}{E_6-E_0} \right].
\label{9}
\end{eqnarray}
\end{full}

Hereafter, we use the abbreviations, $\widetilde{L} \equiv L+D_1+D_2$,
 $\widetilde{h} \equiv h_1-h_2$, $h_1^\pm \equiv h_1\pm
 D_1$, and $h_2^\pm \equiv h_2\pm D_2$. 
In eqs.(\ref{9}), $E_0$, $E_1$, and $E_2$ should be represented 
differently, whether $\widetilde{h}=0$ or not. When $\widetilde{h} = 0$,
\begin{eqnarray}
E_0 & = & 
-\frac{1}{2}\left[\sqrt{(J+\widetilde{L})^2+8J^2}+J+\widetilde{L}\right],
\nonumber \\
E_1 & = & -J-\widetilde{L},
\nonumber \\
E_2 & = & \frac{1}{2}\left[\sqrt{(J+\widetilde{L})^2+8J^2}-J-\widetilde{L}\right].
\end{eqnarray}

\begin{full}
\begin{eqnarray}
&    & \hspace{-4em}\mbox{When $\widetilde{h} \neq 0$,} \nonumber \\
E_n & = & \frac{2}{3}\left[p\cos{\left(\frac{1}{3}\left\{\arccos{\left[p^{-3}(J+\widetilde{L})
 \left\{(J+\widetilde{L})^2+9J^2-9\widetilde{h}^2\right\}\right]}+2\pi (n+1)
\right\}\right)}
 -J-\widetilde{L}\right], \nonumber \\
p & = & \left[(J+\widetilde{L})^2+6J^2+3\widetilde{h}^2\right]^{1/2}, 
\hspace{2em} n=0,1,2.
\end{eqnarray}
\end{full}
\noindent
On the other hand, 
$E_3$,$E_4$, $E_5$, and $E_6$ can be described independently of 
$\widetilde{h}$:
\begin{eqnarray}
E_3 & = & \frac{1}{2}(h_1^-+h_2^--x^-), \nonumber \\
E_4 & = & \frac{1}{2}(h_1^-+h_2^-+x^-), \nonumber \\
E_5 & = & \frac{1}{2}(-h_1^+-h_2^+-x^+),\nonumber \\
E_6 & = & \frac{1}{2}(-h_1^+-h_2^++x^+),
\end{eqnarray}
\noindent
using the abbreviation
$x^\pm=[4J^2+(h_1^\pm-h_2^\pm)^2]^{1/2}$. 
$A_k$ and $B_k$ also depends on $\widetilde{h}$. 
\begin{full}
\begin{eqnarray}
&    & \hspace{-3.5em}\mbox{ When $\widetilde{h} = 0$,} \nonumber \\ 
A_0 & = & B_0 = A_2 = B_2 = 0, \nonumber \\ 
A_1 & = & \sqrt{2}J(K_1+M_1)r_0^{-1}, \nonumber \\
B_1 & = &-\sqrt{2}J(K_2+M_2)r_0^{-1}, \nonumber \\
A_3 & = & [2J^2-(E_0+J+\widetilde{L})(h_1^--h_2^-+x^-)]K_1(\sqrt{2}r_0r_3)^{-1}
, \nonumber \\
B_3 & = & [2(E_0+J+\widetilde{L})-(h_1^--h_2^-+x^-)]JK_2(\sqrt{2}r_0r_3)^{-1},
\nonumber \\
A_4 & = & [2J^2-(E_0+J+\widetilde{L})(h_1^--h_2^--x^-)]K_1(\sqrt{2}r_0r_4)^{-1}
,\nonumber \\
B_4 & = & [2(E_0+J+\widetilde{L})-(h_1^--h_2^--x^-)]JK_2(\sqrt{2}r_0r_4)^{-1},
\nonumber \\
A_5 & = & [2(E_0+J+\widetilde{L})+h_1^+-h_2^+-x^+]JK_1(\sqrt{2}r_0r_5)^{-1},
\nonumber \\ 
B_5 & = & [2J^2+(E_0+J+\widetilde{L})(h_1^+-h_2^+-x^+)]K_2(\sqrt{2}r_0r_5)^{-1},\nonumber \\
A_6 & = & [2(E_0+J+\widetilde{L})+h_1^+-h_2^++x^+]JK_1(\sqrt{2}r_0r_6)^{-1}, 
\nonumber \\
B_6 & = & [2J^2+(E_0+J+\widetilde{L})(h_1^+-h_2^++x^+)]
K_2(\sqrt{2}r_0r_6)^{-1}, \nonumber \\
r_0  & = & [(E_0+J+\widetilde{L})^2+2J^2]^{1/2}.
\label{12}
\\
&   & \nonumber \\
&   & \hspace{-3em} \mbox{And when $\widetilde{h} \neq 0$,} \nonumber \\
A_0 & = & -4\widetilde{h}J^2(E_0+J+\widetilde{L})(K_1+M_1)r_0^{-2}, 
\nonumber \\
B_0 & = & 4\widetilde{h}J^2(E_0+J+\widetilde{L})(K_2+M_2)r_0^{-2}, \nonumber \\
A_1 & = & -2\widetilde{h}J^2(E_0+E_1+2J+2\widetilde{L})(K_1+M_1)(r_0 r_1)^{-1},
\nonumber \\
B_1 & = & 2\widetilde{h}J^2(E_0+E_1+2J+2\widetilde{L})(K_2+M_2)(r_0 r_1)^{-1},
\nonumber \\
A_2 & = & -2\widetilde{h}J^2(E_0+E_2+2J+2\widetilde{L})(K_1+M_1)(r_0 r_2)^{-1},\nonumber \\
B_2 & = & 2\widetilde{h}J^2(E_0+E_2+2J+2\widetilde{L})(K_2+M_2)(r_0 r_2)^{-1},
\nonumber \\
A_3 & = & (E_0+J+\widetilde{L}-\widetilde{h})
[2J^2-(E_0+J+\widetilde{L}+\widetilde{h})
(h_1^--h_2^-+x^-)] K_1(\sqrt{2}r_0r_3)^{-1}, \nonumber \\
B_3 & = & (E_0+J+\widetilde{L}+\widetilde{h})
[2(E_0+J+\widetilde{L}-\widetilde{h})-(h_1^--h_2^-+x^-)]
JK_2(\sqrt{2}r_0r_3)^{-1}, \nonumber \\
A_4 & = & (E_0+J+\widetilde{L}-\widetilde{h})
[2J^2-(E_0+J+\widetilde{L}+\widetilde{h})
(h_1^--h_2^--x^-)] K_1(\sqrt{2}r_0r_4)^{-1},\nonumber \\
B_4 & = & (E_0+J+\widetilde{L}+\widetilde{h})
[2(E_0+J+\widetilde{L}-\widetilde{h})-(h_1^--h_2^--x^-)]
JK_2(\sqrt{2}r_0r_4)^{-1},\nonumber \\
A_5 & = & (E_0+J+\widetilde{L}+\widetilde{h})
[2(E_0+J+\widetilde{L}-\widetilde{h})+h_1^+-h_2^+-x^+
]JK_1(\sqrt{2} r_0r_5)^{-1}, \nonumber \\
B_5 & = & (E_0+J+\widetilde{L}-\widetilde{h}) 
[2J^2+(E_0+J+\widetilde{L}+\widetilde{h})
(h_1^+-h_2^+-x^+)]K_2(\sqrt{2} r_0r_5)^{-1}, \nonumber \\
A_6 & = & (E_0+J+\widetilde{L}+\widetilde{h})
[2(E_0+J+\widetilde{L}-\widetilde{h})+h_1^+-h_2^++x^+
]JK_1(\sqrt{2}r_0r_6)^{-1},\nonumber \\
B_6 & = & (E_0+J+\widetilde{L}-\widetilde{h}) 
[2J^2+(E_0+J+\widetilde{L}+\widetilde{h})
(h_1^+-h_2^++x^+)]K_2(\sqrt{2}r_0r_6)^{-1}, \nonumber \\
r_n & = & [\{(E_n+J+\widetilde{L})^2-\widetilde{h}^2\}^2+
    2\{(E_n+J+\widetilde{L})^2+\widetilde{h}^2\}J^2]^{1/2},\mbox{ } n=0,1,2.
\label{14}
\end{eqnarray}
\end{full}
\noindent
In eqs.(\ref{12}) and (\ref{14}), 
$r_3$, $r_4$, $r_5$, and $r_6$ are common:
\begin{eqnarray}
r_3 & = & [2x^-(x^-+h_1^--h_2^-)]^{1/2},\nonumber \\
r_4 & = & [2x^-(x^--h_1^-+h_2^-)]^{1/2},\nonumber \\
r_5 & = & [2x^+(x^+-h_1^++h_2^+)]^{1/2},\nonumber \\
r_6 & = & [2x^+(x^++h_1^+-h_2^+)]^{1/2}.
\end{eqnarray}

\end{document}